# Effect of disorder on the temperature dependence of radiative lifetimes in V-groove quantum wires

D.Y. Oberli[*], M.-A. Dupertuis, F. Reinhardt[†], and E. Kapon
Département de Physique, Ecole Polytechnique Fédérale de Lausanne, CH-1015 Lausanne, Switzerland

We have studied the effect of disorder on the radiative properties of semiconductor quantum wires by time-resolved photoluminescence spectroscopy. The dependence of the radiative lifetimes is measured over a temperature range extending from 8 to 150 K. At low temperatures we find that the measured dependence does not conform to the theoretical prediction for recombination of free excitons in a quasi one-dimensional system due to the major influence of disorder. An extension of the theory is developped in order to take into account exciton ionization and the contribution of free carriers. A mean localization length for excitons is also estimated from the radiative lifetime at 8 K.

The interaction of light with excitons is profoundly affected by the reduced dimensionality in semiconductor nanostructures. The breaking of translation symmetry in quantum structures leads to the existence of resonant exciton-polaritons, which have finite radiative lifetimes[1]. Whereas the lifetime of free excitons confined in GaAs heterostructures was calculated to lie in the range of tens of picoseconds in ideal quantum wells[2-3] (QWs), it was predicted to increase by an order of magnitude in ideal quantum wires[4] (QWRs). These theoretical predictions assume that excitons are free to move in the unconfined directions and, thus, that only excitons with a wavevector of their center of mass less than $\mathbf{k}_0$ ($\mathbf{k}_0$ is the wavevector of a photon inside the structure with the same energy as the exciton at the zone center) can decay radiatively due to conservation of energy and momentum.

In real heterostructures, however, disorder is present because of spatial fluctuations of the quantum heterostructure size and of the alloy composition. Localization of excitons results from this disorder and leads to enhanced radiative lifetimes[5] because the localized exciton wavefunction in k-space comprises wavevector components larger than $k_0$, which are non-radiative in nature in the polariton picture. These inherent difficulties have been recognized in several experimental studies of excitons in QWs and two main approaches emerged to measure their intrinsic radiative lifetimes (i.e. for free excitons). One study used a very high quality QW sample and was performed at very low temperature under resonant photoexcitation in order to evaluate the radiative lifetime of free excitons[6]. The other studies[7-8-9] relied, instead, on the predicted linear dependence on temperature of the radiative lifetime, which is characterized by a slope proportional to the intrinsic radiative lifetime. The fundamental linear dependence of the radiative decay time with temperature was actually observed under resonant excitation conditions over a narrow temperature range[9]. Disorder effects can similarly alter the intrinsic radiative lifetimes in QWRs. Previous experiments[10-11] reported on the radiative decay of excitons in QWRs and, from the predicted dependence for 1D-excitons as a square root of temperature[4], attempted to estimate the intrinsic radiative lifetimes of free excitons. Although the temperature dependence of the decay times of the QWRs were found to be systematically weaker than those of reference QWs in both experiments, the decay times were essentially the same as in QWs at the lowest temperature (10K).

In this paper, we report a study of the temperature dependence of the photoluminescence (PL) decay times in high quality V-groove QWRs. We show, in particular, that it is important to account for non-radiative recombination in order to estimate the temperature dependence of the radiative lifetimes from the measured PL decay times. We find that the observed temperature dependence is stronger than the predicted one for free 1D excitons and is quasi-exponential over a wide temperature range, which we attribute to the combined effect of disorder and exciton ionization. At the lowest temperature the radiative lifetimes are strongly enhanced by disorder, ranging from 350 to 400 ps, and show almost no dependence on QWR sizes, which indicates the crucial role played by disorder in these studies.

The GaAs-QWRs embedded in $Al_{0.3}Ga_{0.7}As$ were grown by low pressure organometallic chemical vapor deposition on V-grooved (001)-GaAs substrates; more details on the growth and morphology of the samples are given elsewhere[12]. We have studied three samples, each consisting of a 0.5 µm pitch lateral array of single QWRs of different sizes. The nominal GaAs layer thicknesses were 5 nm, 2.5 nm, and 1.5 nm, resulting in a thickness at the crescent center of 14.1 nm, 8.8 nm, and 4.1 nm, respectively. The samples were mounted onto the cold finger of a continuous-flow helium cryostat where temperature can be adjusted between 4.5K and 300K. The luminescence was excited with 3 ps light pulses from a mode-locked titanium-sapphire laser operating at 76 MHz by focusing them onto a 40 µm spot. The emitted light was dispersed by a 46 cm spectrograph and detected with either a cooled CCD for time-integrated PL spectra or a cooled microchannel plate photomultiplier for time-resolved PL. The temporal evolution of the PL was obtained by a time-correlated single photon-counting technique[13] and a typical instrumental response of 60 ps at FWHM was achieved.

In Fig. 1 we display time-integrated PL spectra from the 2.5 nm-QWR sample for temperatures of 10, 100 and 140K. A photon energy of 1.660 eV was chosen in order to selectively excite carriers in the QWR. In these spectra the high energy tail of the substrate luminescence was removed. The untreated PL spectrum at 100 K is shown in the inset. The main spectral features are identified and the position of the higher energy optical transitions are determined from PL excitation (PLE) spectra[14]. The occupation of excited states associated with 1D subbands gives rise to the additional structures on the high energy tail of the QWR-luminescence, which appear if the temperature is raised above about 40 K. Also shown is a standard fit to the high energy tail of the substrate luminescence corresponding to band to band recombination of a thermalized distribution of electrons and holes. This contribution was removed in order to better determine the temperature dependence of the luminescence efficiency of the QWR.

In Fig. 2 we show typical temporal profiles of the QWR-PL[15] from the same sample (QWR-2.5nm) for temperatures of 15, 60 and 100K. The measured profiles are characterized essentially by a mono-exponential decay of the QWR-PL at these and all other temperatures. The average power density was fixed at 10 W/cm$^2$. We verified that at 8K the decay of the profile did not change when the power density was decreased by a tenfold factor. Assuming a 1% absorption we estimate an upper value of the resulting carrier density in the QWR of approximately 5x10$^3$ cm$^{-1}$. At this excitation density excitons in the lowest 1D subband make the dominant contribution to radiative recombination in the QWRs. This is attested to by both the time-integrated PL spectrum (see Fig.1) and the time-resolved spectrum that is measured with a delay time short compared to the decay time (about 100 ps). In this later case, there is no evidence at low temperature for recombination of excitons associated with 1D excited states, which implies that relaxation to the excitonic ground state is more efficient than radiative recombination. We also found that the decay time of the QWR-PL did not depend on the excitation wavelength: The photon energy was progressively decreased in order to coincide with the lowest excitonic resonance in the PLE spectrum, thus reaching resonant excitation of delocalized excitons. Decay times are obtained by fitting the temporal profile to an exponential function over one order of magnitude from the point corresponding to 1/2 the maximum intensity.

The temperature dependence of the decay times is displayed in Fig. 3 for the 2.5 nm-QWR. At the lowest temperatures the decay time is nearly constant and as the temperature is raised above about 20 K it increases rapidly to reach a maximum. An increase of the decay times with temperature indicates that radiative recombination dominates. The observed saturation of the decay time is, however, the signature of competing non-radiative processes that are activated and eventually dominate recombination at elevated temperatures for low excitation density[16]. Indeed, measurement of the temperature dependence on different positions of the same sample yields a similar curve with the same decay time at 8K but with a maximum value that can vary by nearly 50%. Furthermore, we find that the time- and spectrally- integrated intensity of the QWR-PL is nearly constant at temperatures below 15 K and then steadily decreases as the temperature is raised. These observations clearly confirm the significant role played by non-radiative processes in the temporal decay of the PL. The radiative lifetime at a given temperature, $\tau_r(T)$, is estimated from the standard expression for the PL radiative efficiency[17]:

$$\eta(T) = \frac{\tau_d}{\tau_r} = \frac{I(T)}{I_0} \qquad (1)$$

where $\tau_d(T)$ is the PL decay time, $I_0$ and $I(T)$ represent the temporally and spectrally integrated intensities of the QWR-PL at 8 K and temperature T, respectively. The temperature dependence of the radiative lifetime exhibits a striking exponential rise when the lattice temperature is raised above approximately 50 K (see Fig. 3). Current theories for the radiative decay of 1D excitons predict a functional dependence that follows a square root of the excitons temperature under two main assumptions: the existence of free excitons and a rapid thermalization of the exciton population[4]. The first assumption implies that the heterostructure is free of intrinsic and extrinsic defects and, thus, that excitons can propagate freely along the wire axis. The second one requires that excitons following their photoexcitation reach a thermal equilibrium on a time scale short compared to their radiative lifetime. It is thus important to examine whether these assumptions are experimentally verified. We note, however, that a significant fraction of excitons will be ionized as the thermal energy approaches the exciton binding energy and, thus, free carriers will also contribute to the radiative recombination. In comparison to the 3D and 2D cases, the contribution of the free electron-hole pairs to radiative recombination is diminished because the Sommerfeld factor is smaller than unity in the 1D case[18].

We have measured the PL and PL-excitation (PLE) spectra of our three samples as a function of temperature. In this study we find a common behavior, which was previously reported[19]: as the temperature is increased the Stokes shift (e.g. 5.4 meV at 8K, 2.5 nm-QWR) progressively reduces and vanishes at a thermal energy, $k_BT$, that can be significantly larger than its magnitude (T~100K). This vanishing of the Stokes shift reflects the thermally induced redistribution of exciton population from localized states to largely extended states at the maximum of the lowest excitonic resonance in the PLE spectrum. Localization of excitons is caused by interface and alloy disorder, which results in fluctuations of the confining potential along the wire axis and thus in the occurence of local potential minima where excitons are bound. Several studies have recently confirmed this picture with the direct observation of sharp excitonic lines that are attributed to these bound states in the microscopic-PL spectra of narrow GaAs/AlGaAs QWs[20] and of V-groove QWRs[21]. These experimental evidences taken together with recent theoretical studies[22] indicate that the influence of disorder on optical spectra persists up to temperatures for which the mean thermal energy already exceeds the inhomogeneous broadening. The weak temperature dependence of the radiative lifetime at the lowest temperatures, below 10 to 40 K depending on the sample, is attributed to exciton localization. The primary effect of disorder is to increase the radiative lifetime of localized excitons over that of free excitons[5]. Assuming that an exciton localized in a local potential minimum is characterized by a gaussian center-of-mass wavefunction with a spatial extent $l_c$, an expression for the radiative lifetime of a localized exciton in a quantum wire[23] is then obtained as

$$\tau \approx \tau_o \frac{3}{1+\alpha} \sqrt{\pi} (k_o \cdot l_c)^{-1} \qquad (2)$$

where $\tau_o = 2/(\Gamma_L + \Gamma_{T_1} + \Gamma_{T_2})$ is the intrinsic radiative lifetime of excitons, $\Gamma_L$ and $\Gamma_{T_{1,2}}$ are respectively the radiative widths for the longitudinal and the two transverse exciton-polariton modes with zero-wavevector, $k_o$ is the photon wavenumber given by $k_0 = \omega n / c$, n is the index of refraction, $\hbar\omega$ is the exciton energy and $\alpha$ is a factor that depends on the shape of the wire. For a cylindrical wire $\alpha$ is exactly equal to 1/5 and, according to Citrin[5], $\Gamma_L = 8\Gamma_{T_{1(2)}}$. The expression for $\tau$ is only valid for localized excitons that satisfy the following relation: $a_B \ll l_c \ll \lambda_o/n$ where $a_B$ and $\lambda_o$ are respectively the exciton Bohr radius and the photon wavelength. In this case the notion of finite coherence area introduced by Feldmann et al.[7] in the context of QWs finds a microscopic justification as the spatial extent of the excitonic center-of-mass wavefunction[24-25], which can be interpreted as a mean localization length characteristic of the disorder resulting from random size fluctuations. From the low-temperature radiative lifetime of 400 ps we estimate this localization length to be about $l_c$=320 Å based on our calculated value of $\tau_o$ =104 ps and of $\alpha$ = 0.48 for the 2.5 nm-QWR structure.

Above 60 K, we find that different spectral components within the high energy tail of the QWR-PL spectrum are characterized by the same decay time. A comparison of the time-integrated PL spectra at temperatures higher than 60K shows that they are well described by a unique spectral profile times the energy dependent Boltzmann factor for that lattice temperature. These findings provide direct evidences that excitons and free carriers have reached a thermal equilibrium. The radiative recombination time can thus be identified as the radiative lifetime of the thermal population of excitons and free carriers. Consequently, the exciton model describing the temperature dependence of a thermal population of excitons cannot be applied at least in the temperature range T ≥ 60 K.

Another model that includes the contribution of free carriers to radiative recombination has been proposed for carriers confined in quantum well structures[26]. We now outline an extension of this model to the case of quantum wires and discuss its predictions in relation to our experimental data. The model is based on two assumptions: a thermodynamic equilibrium between free carriers and excitons that is described by the Saha equation and a finite density of background holes. In quantum wires, the Saha equation is given by

$$\frac{N_e N_h}{N_X} = \left(\frac{2\mu k_B T}{\pi \hbar^2}\right)^{1/2} \exp\left(-\frac{E_B}{k_B T}\right) \equiv K_e(T) \quad (3)$$

where $N_e$, $N_h$, $N_X$ are, respectively, the density of electrons, holes, and excitons, $\mu$ is the reduced effective mass of the exciton, T is the temperature and $E_b$ is the exciton binding energy. Introducing the total number of photoexcited electron-hole pairs, $N = N_e + N_X$, the fraction of exciton population, $N_X/N$, at a given temperature can be expressed from Eq. (3) as

$$\frac{N_X}{N} = \frac{p_o}{p_o + K_e(T)} \quad (4)$$

in the limit where the background hole density $p_o$ is much larger than N. The total radiative recombination rate in QWR structures with a background p-doping, $1/\tau(T)$, is the sum of two contributions: the first corresponds to the radiative recombination rate of electrons with background holes and the second to that of the effective exciton population at the temperature T. The expression for the temperature dependence of the total rate then reads

$$\frac{1}{\tau(T)} = \left\{\frac{1}{\tau_o^f}\left(\frac{\mu}{(m_e m_h)^{1/2}}\right)\exp\left(-\frac{E_b}{k_B T}\right) + \frac{1}{\tau_o}\frac{4(1+\alpha)}{3}\left(\frac{E_1}{\pi k_B T}\right)^{1/2}\right\}\frac{p_o}{p_o + K_e(T)} \quad (5)$$

where $E_1 = \hbar^2 k_o^2/2M$ is the maximum kinetic energy of radiative excitons, $m_e$, $m_h$, M (M = $m_e + m_h$), are, respectively, the effective masses for electrons, holes, and excitons, $\tau_o$ is the radiative lifetime of excitons and $\tau_o^f$ is the radiative lifetime of free carriers, which is weakly dependent on temperature through the band gap energy (see e.g. Ref. 26). We note that the PL is described by a single exponential decay when the condition for the validity of Eq. (4) is fulfilled; this corresponds to our experimental condition of low excitation density. In Fig. 4 we report a comparison between the prediction of the model and our data for the 2.5 nm-QWR structure. As the temperature is increased beyond 60 K, where a thermal equilibrium is known to exist experimentally, the exciton population decreases steadily (see inset in Fig. 4) and, thus, the radiative rate is reduced by the remaining fraction of excitons at a given temperature. We find that this model can account for the order of magnitude increase of the radiative recombination times assuming a reasonable value for the background density of holes ($p_o = 5 \times 10^4$ cm$^{-1}$), which is the only free parameter[27]. It is worth mentioning that the dependence on temperature is dominated by the reduction of the exciton population that can recombine radiatively. Although the expression for the radiative recombination rate of the free carriers has been simplified by omitting the Sommerfeld factor, which accounts for Coulomb correlations of unbound electron-hole pairs, and by assuming that the overlap integral of electron and hole enveloppe wavefunctions in the confinement directions is one, the predicted temperature dependence is not modified. Nevertheless, the observed temperature dependence of the radiative recombination times is not reproduced by this model. We attribute this different temperature dependence to the persistent role of disorder at elevated temperatures as was already inferred from the studies of the temperature dependence of the Stokes shift.

In order to specify this role we present in Fig. 5 a comparison of the temperature dependence of the radiative lifetimes for our three samples that are characterized by different degrees of disorder. The disorder is assessed by the inhomogeneous broadening as estimated from the full width at half maximum ($\Gamma$) of the first optical transition in a PLE spectrum obtained at 8 K[14]. First to be noted is the similarity of the temperature dependence for the three sets of data; namely, a quasi-exponential rise of the radiative time at temperatures above 50 K is found. The steepness of this rise is, however, distinct and correlated to the nominal thickness of the QWR. This trend can be understood in the framework of the radiative model for thermalized and free excitons: besides the square root of temperature dependence predicted by that model, the lifetime is inversely proportional to the exciton oscillator strength[4]. Therefore the more tightly confined excitons, e.g. in sample 1.5nm-QWR, are expected to have the weakest temperature dependence and also the lowest value of the radiative lifetime at a given temperature. Our results are consistent with this description although excitons in the present QWR-structures cannot be regarded as free excitons. The similar lifetimes observed at the lowest temperature in the three samples are attributed to the effect of localization that enhances more effectively the radiative lifetime of excitons in the presence of stronger disorder (i.e. for larger inhomogeneous broadening) thereby balancing the inherently larger oscillator strengths of thinner wires.

With regard to the temperature dependence of inhomogeneously broadened 2D excitons we remark that it follows the same law[8] indicating that dimensionality of the exciton confinement may not be relevant, but rather that the magnitude of the inhomogeneous broadening of the exciton at low temperature plays a key role. In this latter study the broadening was comparable to the exciton binding energy. On the other hand, when the exciton resonance was much narrower than the binding energy, a quasi-linear temperature dependence of the radiative lifetime was observed in several experiments[7-9] for a narrow temperature range that did not exceed 30-50 K. The radiative lifetime has also been calculated for two-dimensional excitons in the presence of weak disorder, implying that the internal motion of the exciton and its C.O.M. motion can be decoupled. With this assumption and in the limit of high temperature ($k_B T \gg \Gamma$) the radiative lifetime of 2D excitons is found to recover the same linear temperature dependence as that of free excitons[28-29-30].

In conclusion, disorder is shown to affect the temperature dependence of the radiative lifetime over a significant temperature range that extends to values of the thermal energy comparable to the inhomogeneous broadening. In the upper part of the measured temperature range (50 to 150 K), the temperature dependence of the radiative lifetime follows a quasi-exponential law, which contrasts with current theoretical predictions for a thermalized distribution of free excitons and free carriers. We believe that these results are generally relevant for excitons confined in disordered nanostructures and that they should stimulate further the theoretical modeling of excitons and their radiative properties when strong disorder is present.

**Acknowledgment**

Helpful discussions with R. Zimmermann and D. Citrin are acknowledged. This work was supported in part by the Fonds National Suisse de la Recherche Scientifique.

* email: Daniel.Oberli@epfl.ch
†current address: Lucent Technologies, Microelectronics, Reading, PA 19612, USA

Figures:

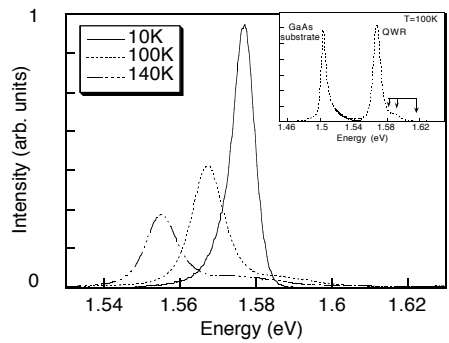

Fig.1: Time-integrated PL spectra showing the QWR line for the sample with a nominal thickness of the GaAs layer of 2.5 nm at temperatures of 10, 100 and 140 K. The spectral shift of the QWR line is mainly caused by the temperature dependence of the GaAs bandgap. Inset shows the full spectrum at 100 K: emission from thermally occupied excited states in the QWR is indicated by arrows; solid line is a fit to the high energy tail of the substrate PL.

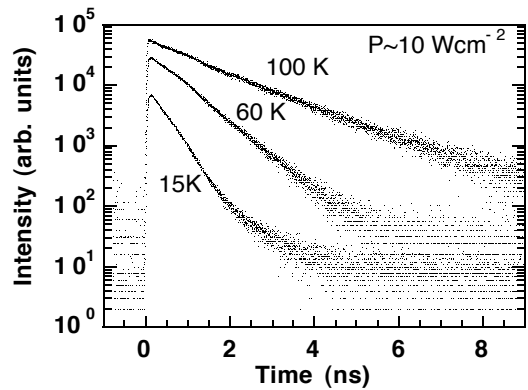

Fig.2: Temporal profiles of the PL emission from the 2.5 nm-QWR at temperatures of 15, 60 and 100K. Curves for 60 and 100K are shifted upwards for clarity.

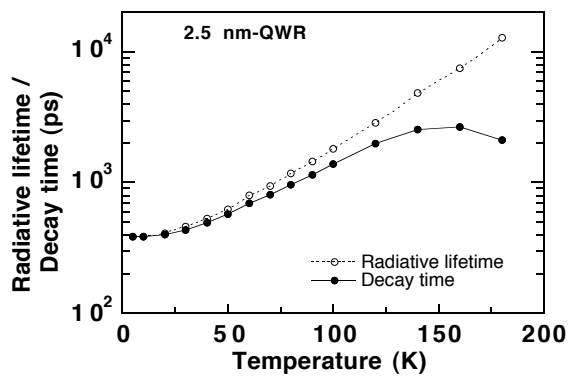

Fig.3: Temperature dependence of PL decay times and radiative lifetimes in the 2.5 nm-QWR sample. Times are displayed on a logarithmic scale; lines are guides to the eyes.

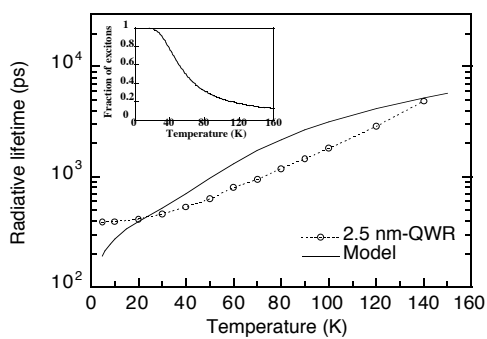

Fig.4: Comparison of the radiative times as obtained for the 2.5 nm-QWR sample with the prediction of the model described by Eq. (5) for the following parameters: $p_o$= $5 \times 10^4$ cm$^{-1}$, $\tau_o^f$= 640 ps, $\tau_o$= 104 ps, $\alpha$= 0.48, $\mu$= 0.048 $m_o$, $m_e$= 0.067 $m_o$, $m_h$= 0.16 $m_o$, $E_1$= 0.1 meV, $E_B$= 11 meV. Inset shows the temperature dependence of the fraction of the exciton population that is calculated with the same parameters.

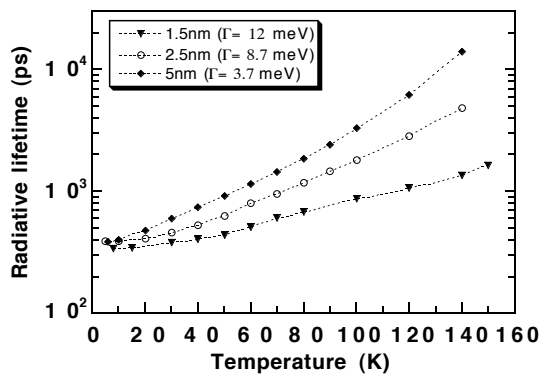

accepted for publication in PRB										9

Fig.5: Temperature dependence of radiative lifetimes for three samples. Different symbols refer to samples with different nominal thicknesses of the GaAs layer (1.5, 2.5 and 5 nm); $\Gamma$ is the inhomogeneous broadening characterizing each sample.